\documentclass[10pt]{iopart}

\newcommand{\tlcf}{two level correlation function}
\newcommand{\pll}{\parallel}
\newcommand{\x}{\perp}
\newcommand{\tErh}{t_{\rm Erh}}
\newcommand{\half}{\textstyle{\frac{1}{2}}}
\newcommand{\third}{\textstyle{\frac{1}{3}}}

\newcommand{\al}{\alpha}
\newcommand{\be}{\beta}
\newcommand{\de}{\delta}
\newcommand{\De}{\Delta}

\newcommand{\om}{\omega}

\newcommand{\brkt}[1]{\left({#1}\right)}
\newcommand{\sqbrkt}[1]{\left[{#1}\right]}
\newcommand{\anbrkt}[1]{\left<{#1}\right>}
\newcommand{\stbrkt}[1]{\left|{#1}\right|}

\input{epsf.sty}
\begin{document}
\bibliographystyle{iopbib}

\title{Can the trace formula describe weak localisation?}
\author{R S Whitney, I V Lerner and R A Smith}
\address{School of Physics and Astronomy, University of Birmingham,
Edgbaston, Birmingham, B15 2TT, United Kingdom.}

\begin{abstract}
We attempt to systematically derive perturbative
quantum corrections to the Berry diagonal approximation of the 
two-level correlation function (TLCF) for chaotic systems.  
To this end, we develop a ``weak diagonal approximation'' 
based on a recent description of the first weak localisation 
correction to conductance in terms of the Gutzwiller trace formula.
This semiclassical method is tested by using it to 
derive the weak localisation corrections to the TLCF for a semiclassically 
disordered system.
Unfortunately the method is unable to correctly reproduce the 
``Hikami boxes'' (the relatively small regions where classical paths
are glued together by quantum processes).
This results in the method failing to reproduce the well known weak 
localisation expansion.  It so happens that for the first order correction 
it merely produces the wrong prefactor.  
However for the second order correction, it 
is unable to reproduce certain contributions, and leads to a result 
which is of a different form to the standard one.  
\end{abstract}

\section{Introduction.}

Much recent research has concentrated on exploiting the similarities between
quantum chaos (the quantum behaviour of systems which are 
classically chaotic) and disordered conductors in order to gain a better
understanding of both [1-7]
% \cite{Argaman1993a,KhM2,Agam1995,bal-sigma,Argaman1995+1996,Aleiner1996,Cohen1998a}.
For both types of system one
cannot  calculate energy levels and eigenfunctions explicitly
since  these do not have
simple analytic forms due to the non-integrability of the systems.
 One therefore
concentrates on the statistical properties of the energy levels and 
wavefunctions.  The key similarity between quantum chaotic and
disordered systems is that, in an appropriate regime, their energy level
spectra both have statistics described by random matrix theory (RMT) [8-10]
% \cite{Efetov1982+1997,Bohigas1984b,Mehta1991}.

The techniques used to analyse the two systems (quantum chaotic and 
disordered conductors) are very different, and it is this which makes 
the similarities in the level statistics so potentially fruitful. 
Quantum chaos makes use of the Gutzwiller trace formula \cite{Gutzwiller1971},
a semiclassical relation in which all classical periodic orbits are summed 
over to obtain the quantum mechanical density of states. 
In most work a particular system is
considered, and statistical averaging is performed over an energy window in 
the spectrum. In disordered conductors,  the
microscopic details of the disorder potential are neither known
nor interesting, and this enables one to average
over all realisation of disorder from the very beginning. This allows one to
develop an effective field theory which has a well-defined perturbation
expansion. As a consequence it is possible to calculate statistical
quantities in disordered conductors in several important regimes. More 
specifically there are two important length scales in a disordered conductor: 
the system size, $L$ and  the elastic transport
mean free path, $\ell=v_F\tau$, where $\tau$ is the elastic transport
scattering time, $v_F$ is the Fermi velocity. Using
the diffusion constant $D=v_F^2\tau/d$, we can generate two time scales:
$\tau$ and the ergodic time, $\tau_{erg}=L^2/D\sim(L/\ell)^2\tau$,
which is the time taken for
an electron to explore the whole sample. This leads to the following time
regimes: the {\it ergodic regime}, $\tau_{erg}<t$, where the electron has
explored the whole sample; the {\it diffusive regime}, $\tau_{erg}<t<\tau$,
where the electron is diffusing through the sample as it
scatters off impurities;
and the {\it ballistic regime}, $t<\tau$, where the electron is moving
ballistically between scatters. These of course lead to related energy regions
via the uncertainty relation: {\it ergodic}, $E<E_{Th}$; 
{\it diffusive}, $E_{Th}<E<\hbar/\tau$; {\it ballistic}, $\hbar/\tau<E$,
where the Thouless energy, $E_{Th}= \hbar/t_{\rm erg}= \hbar D/L^2$.
Finally, there is the energy scale $\Delta$, the mean energy level
spacing which defines a further energy regime within the ergodic regime:
the {\it quantum regime}, $E<\Delta$,
corresponding to times longer than the Heisenberg time, 
$t>t_{\rm H}=\hbar/\Delta$.  In the disordered conductor one can
calculate the level statistics in all these regimes,
\cite{Efetov1982+1997,Altshuler1986,AG}
 and it shows RMT
behaviour in the ergodic regime. In contrast,  
the Gutzwiller approach has so
far been consistently applied within
the diagonal approximation developed by
Berry \cite{Berry1985},
although efforts have been made to go beyond this 
approximation by considering  action correlations\cite{Argaman1993b}. 
In the diagonal approximation, the leading contribution to 
the two-level correlation function
(TLCF) has been found in the ergodic regime of quantum chaotic
systems \cite{Berry1985}, and in the ergodic and
diffusive regime of disordered systems\cite{Argaman1993a}. 
Unfortunately, this method only works outside the quantum
regime; 
in particular, it does not allow one to calculate  
the corrections in powers of $t/t_H$
predicted by RMT in the ergodic regime for systems with time-reversal
invariance.
 This might lead one to ask the following questions: (i) %
 Are
the level statistics in quantum chaotic systems still described by
RMT within the quantum
regime? In particular, do the corrections in powers of $t/t_H$
predicted by RMT exist in the ergodic regime? 
(ii) Are there any analogues of the diffusive or ballistic regimes
in quantum chaotic systems?

 It is the possibility of answering such questions
as these that motivates work into the analogies between chaotic and disordered
systems.

One approach to answering such questions would be to develop a powerful
field theoretical technique for quantum chaotic systems analogous to that
which exists for disordered metals\cite{Wegner1979,sigma}.
This field theory should be able to
reproduce Berry's result as the first term in a perturbation expansion.
An effective field theory -- the ``ballistic'' sigma-model [2-4]
% \cite{KhM2,Agam1995,bal-sigma} 
-- has recently been developed for quantum chaos. The only averaging used was
over an energy window in the spectrum, and this seemed sufficient to reproduce
RMT behaviour. However,
this method lacks a well-defined perturbation expansion,
unless one introduces additional statistical averaging over some
ensemble [18-20]
% \cite{Zirn1998,Tripani1998,Blanter1998}. 

The aim of this paper is to establish whether a regular perturbative 
expansion can be formulated within semiclassical methods based on the
Gutzwiller trace formula. 
In particular, we will concentrate on the two-level correlation
function (TLCF), $R(\omega)$, and its Fourier transform, the spectral form
factor (SFF), $K(t)$.  We consider the model of randomly
distributed semiclassical scatterers\cite{Argaman1995+1996,Aleiner1996}.
Standard scaling considerations
\cite{Abrahams1979} ensure that for sufficiently large systems the results 
should coincide with those derived for a system with Gaussian
white-noise disorder \cite{Altshuler1986,Kravtsov1995}, which is the 
conventional model for diagrammatic considerations.   
Note that in the diffusive regime in 
two dimensions  the one loop
result (corresponding to the diagonal approximation)
vanishes, and the two loop result is the leading contribution
\cite{Kravtsov1995}
 for
systems with time-reversal invariance, whilst the three loop result is the
leading contribution for systems without time-reversal invariance. 
 
Verifying such an approach for disordered systems should be
the first step before
one can apply the same methods to generic quantum chaotic
systems. Perturbation theory is used to describe the 
diffusive regime, whose  existence in the disordered system is due 
to the separation of the time scales, $\tau$ and $t_{\rm{erg}}$, by
the large parameter $(L/\ell)^2$. Such a parameter does not exist in
a generic chaotic system. Nevertheless, numerically $t_{\rm{erg}}$ 
is always much longer than the `ballistic' time, $\tau$.  
It might be that the perturbation expansion of $K(t)$ for
 $\tau<t<t_{\rm{erg}}$ would reveal behaviour which
is more universal than that characterised just by the contribution 
from short orbits.  
Perturbation theory is also used to describe the ergodic regime of disordered 
systems.  It is known from the exact solution \cite{Efetov1982+1997} that 
there are 
correction in powers of $t/t_H$ to the leading order diagonal contribution to 
the SFF (which is linear in $t$).  For such systems,
perturbation theory is able to reproduce these corrections.
However in the case of a generic chaotic system, it is not known whether such 
corrections necessarily exist; if perturbation theory could be applied to 
these systems this could be checked.

 Therefore, we will try to develop a regular perturbation expansion
based on the Gutzwiller approach.
In this case, the SFF is
contributed by the sum over all pairs of closed orbits which particles traverse
in the system. The diagonal approximation \cite{Berry1985}
suggests that only
identical (or time-reversed, if allowed) paths make a contribution which does
not disappear after averaging over some energy interval. The reasoning behind
this is that different orbits have uncorrelated classical actions, and hence
their contributions to the SFF have uncorrelated phases and average to zero.

Perturbation theory for $K(t)$ in the disordered 
metal \cite{Altshuler1986,Argaman1993a} 
gives not only  the diagonal result in leading
order, but also weak localisation corrections \cite{Kravtsov1995}
 at higher order.
Recently we have shown\cite{Smith1998} how to interpret the
weak localisation corrections  for disordered  metals in terms of  
 a trajectory picture.
It is tempting to use this trajectory picture for actual calculations 
in the language of the trace formula. 

In the trajectory picture, we identified paths contributing
to the SFF which are identical for most of their length, except at regions where
a path self-intersects. At these regions it is possible to make two different 
paths by connecting up the partial paths in different ways. We thus obtain a
set of contributions to the SFF where the actions corresponding to the two paths
are almost identical, whilst the paths themselves are piecewise identical
(see Fig.\ \ref{fig10}).  We
introduce the terminology {\it weak diagonal approximation} to describe such
paths in contrast to Berry's {\it strong diagonal approximation}. The
perturbative order of a given contribution is the number of loops created
by self-intersections
(so that the strong diagonal approximation is the one-loop term). 
In disordered metals, the
 regions where the paths are identical correspond
to diffusion propagators;
the self-intersection regions to the Hikami 
boxes \cite{Hik:81,sigma}. 
The latter are regions
with size of order $\ell$ where $s$-wave scattering ensures the
gluing together of diffusive paths. In a general chaotic system, the loops 
are made by gluing
together  (by diffraction, tunneling or some other quantum process)
 classical trajectories to make
a closed orbit. By analogy with the disordered metal, we call such a  gluing
region a Hikami box. We
believe that the whole question of whether a semiclassical approach can
reproduce perturbation theory will reduce to the question of whether it can
correctly obtain the Hikami boxes -- they are the glue which holds the coherent
partial paths together.

To carry out the test of semiclassical theory proposed in the last paragraph,
we need to use a particular semiclassical approach and work up to a particular
order of perturbation theory. 
A number of semiclassical methods exist for dealing with what we call the 
Hikami boxes [25-27,5]
% \cite{LKh82+Kh84,Bergman1984,Chakravarty1986,Argaman1995+1996}.
However since these only calculate the leading order weak localisation 
correction to conductivity in disordered metals, they deal with a relatively 
simple Hikami box.
To test the validity of a semiclassical approach 
 one should check whether it reproduces the higher order 
Hikami boxes.
We take the approach based on the Gutzwiller trace formula, developed by 
Argaman\cite{Argaman1995+1996}, as such an approach would have a good 
chance of being directly applicable to chaotic systems.
  We calculate the SFF, and hence the TLCF,
to third loop order in the perturbation expansion (which involves
the calculation of higher order Hikami boxes).  Note that the two (three)
loop order gives the leading
contribution to the TLCF in the diffusive regime of
 a disordered system with (without) time-reversal symmetry.
Unfortunately, this  semiclassical method gives the correct 
functional form only up to two loop order. 
Beyond this, it does not work. In particular,
we show that it can not reproduce the three-loop result.
It appears that this method, although quite general and 
very attractive because of
its simplicity and obvious physical interpretation, cannot correctly
obtain the Hikami boxes.
This does not mean that the Hikami boxes cannot be 
reproduced by semiclassical methods: it has been shown 
\cite{Aleiner1996} that this is possible for exactly the system we 
consider in this paper.  However, it is not clear whether the technique
suggested in \cite{Aleiner1996}  may be extended to generic chaotic
systems.  Since the method we consider here is based on the Gutzwiller 
formula, if it were successful, it would be completely general.

The rest of the paper is organised as follows. 
In section 2, we discuss the weak diagonal approximation and the regimes in 
which it will apply. In section 3, we briefly review 
the derivation, from the Gutzwiller trace formula, of a semiclassical 
expression for the SFF, and the application of the strong diagonal 
approximation to this expression for the SFF.  We then construct the
weak diagonal expansion in terms of this semiclassical expression for the SFF.
In sections 4 and 5 we calculate
the two-loop and three-loop contributions to the SFF (the first and second
terms in the weak diagonal expansion) respectively. In sections 6 and 7 we
use the results of sections 4 and 5 to write explicit formulas for the
leading contributions to the TLCF of a two-dimensional disordered system
in the diffusive regime with and without time-reversal symmetry. In section 8
we consider whether the techniques used for the diffusive regime can be 
extended to the ergodic regime. Finally in section 9 we discuss our results.

%-----------------------------------------------
\section{The Weak Diagonal Approximation.}

In this section we explain in more detail what is meant by the 
weak diagonal approximation. 
We then discuss the timescales for which such an approximation would 
be valid.

\begin{figure}
\centerline{\epsfbox{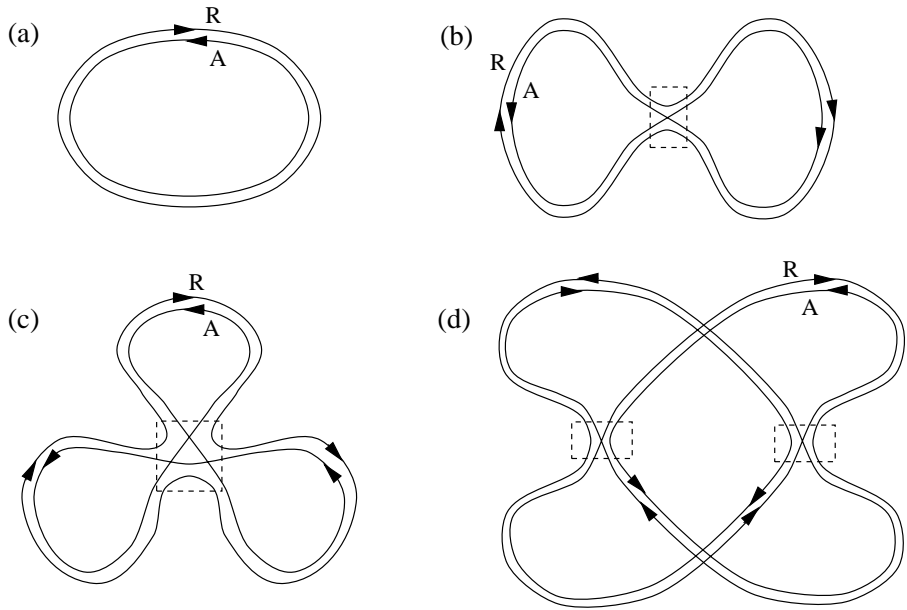}}
\caption{A schematic representation of the first few contributions to
the weak diagonal approximation of the TLCF.  The two paths in each figure 
follow each other exactly, except in the dashed boxes.  In the dashed boxes,
which correspond to Hikami boxes, 
quantum scattering causes one path to cross while the other does not. 
In (a) the paths are identical everywhere, this gives  $R_1(\om)$
which equals the strong diagonal approximation.  The geometry of paths 
contributing to $R_2(\om)$ are shown in (b).  The two different geometries of 
paths contributing to $R_3(\om)$ in a system without time reversal symmetry, 
are shown in (c) and (d). 

The arrows on the retarded Green's functions (R) point in the direction of the 
underlying path, while those on the advanced Green's functions (A) point in 
the opposite direction.}
\label{fig10}
\vspace*{0.1in}
\end{figure}

The weak diagonal approximation involves evaluating the contribution of paths 
that are {\it nearly-identical}.  In this context {\it nearly-identical} 
means the pair of paths follow each other everywhere, except
when the paths come {\it close} to themselves. In such regions (``Hikami
boxes'' marked by dashed boxes in 
Fig.\ \ref{fig10}) the two paths differ:  
one path crosses in this region, while the other does not.
Two paths are considered  {\it close} 
when their separation in phase space satisfies $\de r \de p \leq \hbar$. 
If the system has time-reversal symmetry 
then $\de p$ is either the sum or the difference of the momenta.

%%%%%%%%%%%%%%%%%%%%%%%%%%%%%%%%%%%%%%%%%%%%
%%%%%% figure 1 (whitfig10.eps) goes here %%%%%%
%%%%%%%%%%%%%%%%%%%%%%%%%%%%%%%%%%%%%%%%%%%%
%%%%%%%%%%%%%%%%%%%%%%%%%%%%%%%%%%%%%%%%%%%%

We consider systems in which the Hikami boxes are due to 
quantum mechanical effects such as diffraction, tunnelling, etc.
The time scale,  $t_Q$,
of such quantum scattering events, 
 will be infinite in the classical limit $\hbar \to 0$, 
and much larger than the flight time (mean free time)
$\tau$ in semiclassical (disordered) systems.   
Nonetheless, these quantum events are of great importance 
for any paths with a period larger than $t_Q$.
These quantum scattering events cause a particle to be 
spread among a bunch of classical paths which are close
in phase space.

For paths with a period much larger than $t_Q$, the contribution of a 
given pair of {\it nearly-identical} paths will be of the 
same order as the contribution of identical paths.
However the probability of a path coming 
close to itself in phase-space  in  time $t$ 
 is small if $t \ll t_{\rm H}$, and hence the weak diagonal 
approximation provides perturbative corrections to 
the strong diagonal approximation.  The strong diagonal 
approximation arises from a path which forms a single loop (Fig.~1a).
The leading order correction comes from the paths with one 
self-intersection which form two loops, as shown in Fig.\ \ref{fig10}b.
The next order correction  comes from the paths with two
self-intersection which form three loops (e.g. Fig.\ \ref{fig10}c,d) and so on. 
This loop expansion is totally equivalent
to the standard field-theoretical loop expansion\cite{Smith1998}. 

 In the ergodic regime, $t_{\rm erg} \ll t \ll t_{\rm H}$, 
the probability  of a path coming close to itself is
of order $t/t_{\rm H}$.  Here one expects universal 
behaviour which is described by RMT. The strong 
diagonal approximation reproduces only the linear behaviour of the form 
factor $K(t)$. The weak 
diagonal approximation allows the calculation of perturbative 
corrections in $t/t_{\rm H}$, thus raising the possibility of checking
whether chaotic systems fully obey RMT \cite{Efetov1982+1997,Bohigas1984b}
in this regime. 
Note that the very possibility of this expansion depends on whether
the condition $t_Q\ll t_{\rm H}$ is satisfied.
This is not necessarily satisfied for a generic billiard of typical 
size $L$, for which
$t_Q \sim\tau (L / \lambda_E )^\alpha $, where the flight time
 $\tau\propto L$, and $\lambda_E$ is the De Broglie wavelength of a particle
with energy $E$. As $t_{\rm H}\propto L^2$ for a two-dimensional billiard,
the above inequality is parametrically valid only for $\alpha <1$.

We have chosen to apply the weak 
the weak diagonal approximation to a system of
 randomly placed semiclassical scatterers because the 
above conditions are easily satisfied.
In this case, the quantum scatterings are
mainly due to the paths grazing the scatterers, so the 
theory of penumbra diffraction \cite{Primack1996+1997}
can be applied.  Then for a system
with  $N$ scatterers of radius $R$ per unit volume 
$t_Q \sim \tau (R / \lambda_E )^{2/3}$, where
 $\tau = (v_E N R^{d-1})^{-1}$.
This guarantees that $t_Q \ll t_{\rm H}$ even if the system
size $L$ is of order $R$.
If we consider an extended system of size $L \gg v_E \tau$, then 
$t_Q \ll t_{\rm erg}$, and the weak diagonal approximation can be 
applied to the diffusive regime of this system.
In this regime the 
probability of self-intersection is parametrically smaller than
 $t/t_{\rm H}$. The quantum parts of trajectories (Hikami boxes)
are the same as in the ergodic regime whilst the classical parts give
a different contribution as they are too short to explore all the phase
space. 
When applied to the diffusive regime of a system with a Gaussian 
white-noise disorder potential, the weak diagonal approximation still
reproduces \cite{Smith1998} the standard weak localisation corrections
\cite{sigma,Altshuler1986}.
The system of semiclassical scatterers in the limit of 
$L\gg R$ must have weak localisation corrections of the same form as those
for the white-noise disorder; therefore, any consistent semiclassical method
must reproduce these results. 

It is interesting to ask whether an 
analogous non-ergodic regime exists for generic chaotic systems. For this 
the time-scales must be arranged as $\tau\ll t_Q\ll
t_{\rm erg}$. Since $\tau$ and $t_{\rm erg}$ are parametrically the
same, the numerical window between them should be large enough to
squeeze $t_Q$ inside. It remains to be seen whether such a condition 
may be satisfied by any billiard.

The final timescale we should consider is the Erhenfest time 
$\tErh= \lambda^{-1} \ln (R /\lambda_E)$,
where $\lambda$ in the Lyapunov exponent \cite{Aleiner1996}.
For a system of semiclassical scatterers it is clear that $\tErh \ll t_Q$
For a generic billiard, $\tErh$ will depend on the logarithm of 
$(L / \lambda_E )$, while it seems likely $t_Q$ will depend on some 
power of $(L / \lambda_E )$.  
In the semiclassical limit, $\lambda_E \ll R$, we expect that such a 
system will also have $\tau \ll \tErh \ll t_{\rm Q}$.
This will be of relevance to the details of how we construct the Hikami boxes
in the next section.

\section{A semiclassical description of the strong and weak diagonal 
approximations.}

The {\tlcf} (TLCF) is given by,
\begin{equation}
R(\omega)={1 \over \nu^2} \Biggl<{\nu \brkt{E + \half \om}
\nu \brkt{E - \half \om} }\Biggr> -1
\label{mid10}
\end{equation}
where $\nu(E)$ is the density of states per unit volume.  Given the spectrum 
of the system, $\{E_n\}$, $\nu (E)$ can be written as,
\begin{equation}
\nu (E) = \frac{1}{L^d} \sum_n \delta \brkt{E-E_n}
\label{mid20}
\end{equation}
and the average density of states, 
$\nu = \anbrkt{\nu (E)}= \brkt{\Delta L^d}^{-1}$.  The averaging denoted by 
$\anbrkt{\cdots}$ is carried out over a certain energy window in the 
spectrum 
of the chaotic system's spectrum.  In disordered systems it is also possible to
average over the ensemble of all realisations of the disorder.   
Ensemble averaging is used in the vast majority 
of work on disordered systems, because it is simpler and better understood. 
In both chaotic and disordered systems, the mean level spacing is given by the 
Weyl rule, so a system of 
volume, $L^d$, at an energy, $E$, has $\De = h^d / \Omega_E$.  The 
volume of the phase space constant energy 
surface is $\Omega_E = (S_d p_E^{d-1} L^d)/v_E$, where $S_d$ is the surface
area of a $d$-dimensional sphere of unit radius, and $p_E$ and $v_E$ are the 
momentum and velocity of a particle with energy, $E$.

We consider a system of non-interacting electrons in a 
potential which varies slowly on the scale of the wavelength of the electrons 
at the Fermi surface. 
In this semiclassical limit, the density of states is given by the 
Gutzwiller trace formula \cite{Gutzwiller1971}, 
\begin{equation}
\fl \nu(E)-\nu=\sum_\alpha \bigg( A_\al (E) \exp \sqbrkt{iS_\alpha(E) / \hbar}
- A^*_\al (E) \exp \sqbrkt{-iS_\alpha(E) / \hbar} \bigg)
\label{mid40}
\end{equation}
where the summation is over all periodic classical paths, $A_\al (E)$ and 
$S_\al (E)$ are the amplitude and action of the $\al$th classical path.
For convenience we have defined $A_\al (E)$ so that it includes the phase 
factor due to the Maslov index.
When studying spectral statistics semiclassically, it is often more convenient
to consider the  spectral form factor (SFF), $K(t)$, than the TLCF itself, 
\begin{equation}
K(t)= \anbrkt{\ \sum_{\al,\be} A_\al A^*_\be \exp \sqbrkt{i(S_\al-S_\be)/ \hbar} 
\delta \sqbrkt{t-\half \brkt{T_\al + T_\be}} \ }
\label{mid60}
\end{equation}
where the period of the classical periodic path, $\al$, is 
$T_\al = {{\rm d}S_\al(E)}/{{\rm d}E}$.
The TLCF is related to the SFF in the limit that 
$\om \ll E$, by
\begin{equation}
R(\omega)= 4 \De^2 \Re \e \sqbrkt{\int_o^\infty \rmd t \ 
\exp \sqbrkt{i \om t/ \hbar}K(t)}
\label{mid50}
\end{equation}

Berry introduced the diagonal approximation of \eref{mid60} \cite{Berry1985},
arguing that after the averaging was carried out, only terms where
$\al = \be$ would still contribute to the summation.  All terms with
$S_\al \neq S_\be$ oscillate wildly as the averaging is carried out, 
and so these terms will be negligible.  Berry's diagonal approximation assumes
that those terms for which $\al \neq \be$ but $S_\al = S_\be$ are rare enough 
that one can ignore them.  We will call this approximation the {\it strong 
diagonal approximation}, to distinguish it from the {\it weak diagonal 
approximation} that we will introduce below. 
When applied to the ergodic regime in a generic chaotic system, the strong 
diagonal approximation was found to reproduce the leading order 
behaviour of the spectral form factor in random matrix theory.
More generally, it was shown that the contribution of these strong diagonal 
terms to the SFF is given by \cite{Argaman1993a}, 
\begin{equation}
K_1(t)= {2 \over \be}{t \over (2 \pi \hbar)^2}P(t)
\label{mid70}
\end{equation}
where $\be = 1$ if the system has time reversal symmetry, and $\be = 2$ if it 
does not.
Here $P(t)$ is the probability of returning to the same point in phase space,
in a time, $t$, integrated over the phase space surface of constant energy, 
$E$.  We now introduce notation that will be of use later on, let 
$f_E(\bi{r}',\bi{p}',t;\bi{r},\bi{p})$ be the probability that a 
particle which is initially at $(\bi{r},\bi{p})$, is at 
$(\bi{r}',\bi{p}')$ after a time, $t$, given that the particle's final 
energy, $E$, is constrained to equal its initial energy.  Then \eref{mid70}
becomes
\begin{equation}
K_1(t) = {2 \over \be}{t \over (2 \pi \hbar)^2}
\int \rmd \bi{r} \rmd \bi{p}_E \ 
f_E \brkt{\bi{r},\bi{p},t;\bi{r},\bi{p}}
\label{mid72}
\end{equation}
where $\rmd \bi{p}_E = \rmd \bi{p} \de \sqbrkt{H(\bi{r},\bi{p})-E}$, 
in other words the integral is restricted to the surface in phase space with 
energy, $E$.  When applied to the diffusive regime of a two-dimensional 
disordered system, this result is $t$-independent, which means it gives no 
contribution to the TLCF.

We now attempt to develop the weak diagonal approximation for a 
semiclassical system based on the Gutzwiller Trace formula. 
To do this we must consider how to deal with the regions where the paths are 
not identical (Hikami boxes).
A path $\al$ that comes close to itself, will typically stay close to 
itself for a time of the order of 
the Erhenfest time  $\tErh$.  Since $\tErh \ll t_{\rm Q}$, 
the nearest quantum scatterer on the path will be some distance from the 
region where the path comes close to itself.  Therefore the behaviour of 
path $\al$ in the Hikami box can be considered as 
classical.  Path $\be$ is the path that will cross in the Hikami box, 
it will be 
identical to $\al$ up to the last quantum scattering before the region in 
which $\al$ is close to itself.  Path $\be$ must leave that scatterer with a
 slightly different momentum from path $\al$, and so moves away from it slowly 
before converging slowly towards the other leg of path $\al$.  Finally $\be$ 
becomes close enough to $\al$ that at the next quantum scattering the two 
paths become identical again.
When path $\al$ comes close to itself, the distance between the two parts of 
the path is of order of a wavelength, $\lambda$.  Path $\be$ is 
never more than that distance away from one or the other of the legs of path 
$\al$.  These distances are much smaller than the lengthscale of the 
potential, $R$.  Therefore the scatterings of path $\be$ are all correlated to
scatterings on one or other leg of path $\al$.  This means that ensemble
averaging will not destroy the contribution of this pair of paths. 
To evaluate the contribution of pairs of paths of this type we construct a 
method based on that used by Argaman \cite{Argaman1995+1996}. 
Since we are assuming that for every path $\al$, there is a path $\be$,
where $\al$ and $\be$ are described above, we can assume the amplitude of path 
$\be$ is approximately equal to that of path $\al$.  
However path $\be$ has to go slightly further than path $\al$, because it 
crosses itself.  This means we expect the 
action of path $\be$ to be slightly greater than that of path $\al$ 
\cite{Argaman1995+1996}.

%%%%%%%%%%%%%%%%%%%%%%%%%%%%%%%%%%%%%%%%%%%%
%%%%%% figure 2 (whitfig20.eps) goes here %%%%%%
%%%%%%%%%%%%%%%%%%%%%%%%%%%%%%%%%%%%%%%%%%%%
\begin{figure}
\centerline{\epsfbox{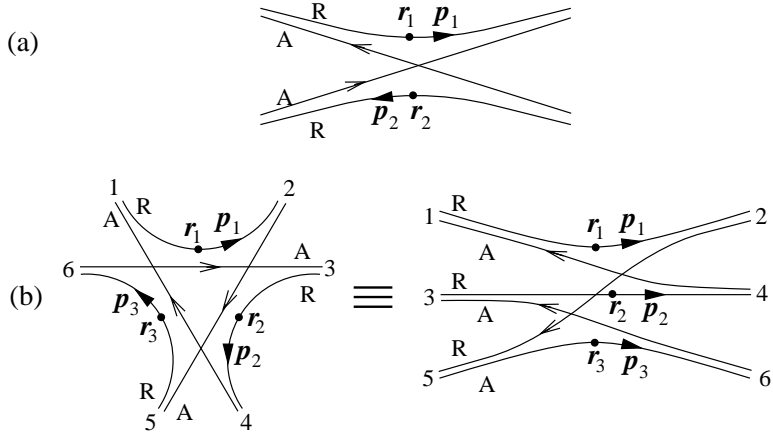}}
\caption{The Hikami boxes (the 
dashed boxes in Fig.\ \ref{fig10}) are shown here in more detail.  In (a) 
$\bi{r}_1$ 
and $\bi{r}_2$ are close, while $\bi{p}_1$ and $\bi{p}_2$ are approximately 
anti-parallel.  In (b)  $\bi{r}_1$, $\bi{r}_2$ and $\bi{r}_3$ are close, 
while $\bi{p}_1$, $\bi{p}_2$ and $\bi{p}_3$ are approximately parallel.
The directions of the arrows are explained in the caption to 
Fig.\ \ref{fig10}.
}
\label{fig20}
\vspace*{0.1in}
\end{figure}
%%%%%%%%%%%%%%%%%%%%%%%%%%%%%%%%%%%%%%%%%%%%

Note that the Hikami boxes in the standard diagrammatic approach
are more complicated: they are `dressed' by additional impurity lines, which
is equivalent to having an extra quantum scattering inside the box. It
is not clear why the contribution of such  an extra scatterer should be
important in the semiclassical approach, since its inclusion would be 
a small correction to the semiclassical Hikami box described above. 
However, the two methods are so different that such a direct comparison
does not make much sense.

We now make the weak diagonal approximation, by assuming we can ignore all 
terms in the double summation in \eref{mid60} which do not have the geometry 
described above.  Clearly terms in the double sum with $\be=\al$ have already
been taken into account by the strong diagonal approximation.
We can then substitute the following into \eref{mid60},
\begin{eqnarray}
A_\be = A_\al \qquad S_\be = S_\al + \de S
\label{mid80}
\end{eqnarray}
Since we are considering pairs of paths that are not identical, the double 
summation over $\al$ and $\be$ in \eref{mid60} gives two contributions for 
each pair of paths.  
For example if we label a given pair of paths, 1 and 2, then
one gets a contribution to the double summation for $\al=1$ and $\be=2$, and 
an identical contribution from $\al=2$ and $\be=1$.
The assumptions in \eref{mid80} allow us to write the $i$th term in the weak 
diagonal expansion of the spectral form factor as,
\begin{eqnarray}
K_i(t) = 2\sum_{\al_i}
\biggl< \stbrkt{A_{\al_i}}^2 \de \brkt{t-T_{\al_i}} 
\exp \sqbrkt{-i \de S_i /\hbar}
\biggr>
\label{mid90}
\end{eqnarray}
where the sum over $\al_i$ is over all periodic paths have $i$ loops.
As can be seen in Fig.\ \ref{fig10}, to derive $K_2(t)$ and $K_3(t)$ we need 
the action difference in two geometries of Hikami box, 
these geometries are shown in Fig.\ \ref{fig20}.  The first is one in which
each path comes close to itself once (Fig.\ \ref{fig20}a) 
and the second is one in which each path comes close to a single point 
on itself twice (Fig.\ \ref{fig20}b). 
In the first case, each path enters the Hikami box twice, 
and the action difference between the two paths is, 
\begin{eqnarray}
\de S_2 = (\bi{r}_1-\bi{r}_2)\cdot(\bi{p}_2+\bi{p}_1)
\label{mid95}
\end{eqnarray}
In the second case, each path enters the Hikami box three times, the 
action difference is, 
\begin{eqnarray}
\de S_{3a} = (\bi{r}_1-\bi{r}_3).\bi{p}_1 + (\bi{r}_2-\bi{r}_1).\bi{p}_2 + 
(\bi{r}_3-\bi{r}_2).\bi{p}_3
\label{mid100}
\end{eqnarray}
The reason for the notation we have chosen for the subscripts on $\de S_2$
and $\de S_{3a}$ will be made clear later in this paper.

%----------------------------------------------
\section{The two loop contribution in the weak diagonal expansion.}

Here we calculate two loop term in the weak diagonal expansion.
This is the leading order correction to the strong diagonal approximation 
(1 loop) result.
The geometry of paths that contribute to the two loop term is shown 
schematically in Fig.\ \ref{fig10}b.  We expect this to be the dominant 
contribution to the weak diagonal perturbation expansion 
in a system with time-reversal symmetry.  However in a system 
without time-reversal symmetry the two loop term in the weak diagonal 
expansion is zero.  This is because one path follows 
the time-reverse of the other path for some of its period.  In a system 
without time-reversal symmetry it is impossible for a path to exactly follow 
the time-reverse of another path.  

In all the figures in this paper we have followed the Feynman diagram 
convention for the arrows.  
This means that the arrows on the retarded Green's functions (R) 
point in the direction of the underlying classical path,
while the arrows on the advanced Green's functions (A) point in the 
opposite direction. This leads to the slightly counter intuitive observation 
that those pairs of paths which have the arrows pointing 
in {\it opposite} directions at all points are the only ones to survive in 
a system without time-reversal symmetry.  In the language of disordered 
systems, the pairs of paths which have the arrows pointing 
in different directions are diffusons, while those with the arrows pointing 
in the same direction are Cooperons.

The two loop term in the weak diagonal expansion,
shown schematically in Fig.\ \ref{fig10}b.  
can be found by substituting \eref{mid95} into \eref{mid90},
\begin{eqnarray}
K_2(t) = 2\sum_\al
\biggl< \stbrkt{A_\alpha}^2 \de \brkt{t-T_\alpha} 
\exp \sqbrkt{-{i\over \hbar}(\bi{r}_1-\bi{r}_2)\cdot(\bi{p}_2+\bi{p}_1) }
\biggr>
\label{1storder110}
\end{eqnarray}
where the $\alpha$th path is a path which starts at 
$(\bi{r},\bi{p})$, passes through 
$(\bi{r}_1,\bi{p}_1)$ at time $t_1$, and $(\bi{r}_2,\bi{p}_2)$ at time $t_2$, 
before returning to $(\bi{r},\bi{p})$ at time, $t$.  
The sum over $\al$ is over all primitive periodic paths and all repetitions 
of them, however in what follows we will ignore the possibility of repetitions.
This is justified in \cite{Argaman1993a} by noting that the number of 
primitive periodic paths grows 
exponentially with the period of the path.  Therefore repetitions make a very 
small contribution to the sum over all paths with a given period.  

The amplitude $A_\alpha$ is the same as in the 
derivation of $K_1(E,t)$, the following result was 
derived in \cite{Argaman1993a},
\begin{eqnarray}
\stbrkt{A_\alpha}^2 \de \brkt{t-T_\alpha}
= {t \over (2 \pi \hbar)^2} \int \rmd \bi{r} \rmd \bi{p}_E \ 
f_{\al E}(\bi{r},\bi{p},t;\bi{r},\bi{p})
\label{1storder120}
\end{eqnarray}
where $f_{\al E}(\bi{r}',\bi{p}',t;\bi{r},\bi{p})$ is the probability of a 
particle following path $\al$ from $(\bi{r},\bi{p})$ to 
$(\bi{r}',\bi{p}')$ in a time $t$, given that the particle's final 
energy, $E$, is constrained to equal its initial energy.  
The integral over momentum is restricted to the energy surface, 
$\rmd \bi{p}_E$ is defined below \eref{mid72}.
Equation \eref{1storder120} was originally derived for purely classical
paths.  We want to consider paths which are mainly 
classical, but which occasionally undergo a quantum scattering.  
For purely classical paths $f_{\al E}(\bi{r}',\bi{p}',t;\bi{r},\bi{p})$ 
is simply a $\de$-function at the 
point in phase space which a classical path starting at $(\bi{r},\bi{p})$ 
would reach after time, $t$.   We include the occasional quantum scattering 
event by dividing a given path up into segments of classical and quantum 
behaviour.  The classical segments will provide most of the path, but there 
will be 
occasional short quantum segments.  The classical segments can be related to 
the classical probability of propagating between two points in a given time.
The quantum segments of the path have to be dealt with by finding the 
quantum mechanical propagator for that scattering event.  If a quantum 
segment of the path goes from $(\bi{r}_i,\bi{p}_i)$ to 
$(\bi{r}_{i+1},\bi{p}_{i+1})$, this can be written in the 
form $A \exp [iR]$ where $A$ and $R$ are defined by the quantum mechanics of 
the scattering event, and are functions of $(\bi{r}_{i,i+1},\bi{p}_{i,i+1})$.
An example of such a quantum scattering event is the penumbra diffraction of
paths that graze semiclassically large scatterers, discussed in 
\cite{Primack1996+1997}.  
The probability of propagating from $(\bi{r}_i,\bi{p}_i)$ to 
$(\bi{r}_{i+1},\bi{p}_{i+1})$ is simply $|A|^2$.  Since the scattering is not 
classical, this probability will not be a $\de$-function.  So a particle on a 
classical path encountering this quantum scatterer  will be spread over a 
number of classical paths after leaving the quantum scattering region.
The probabilities of propagating along each segment of the path can be 
multiplied together to give the probability of following the whole of 
that path.  This procedure gives a result of the same form as 
\eref{1storder120}, however the probability 
$f_{\al E}(\bi{r}',\bi{p}',t;\bi{r},\bi{p})$ can now be written down for 
any path, whether it is purely classical or includes quantum scattering 
events, such as diffraction, tunnelling, etc.
Similarly, when the sum over $\al$ is carried out it is now
a sum over all the paths including quantum scattering events rather than just 
purely classical paths. 
In the limit of paths much longer than $t_Q$, a particle can be considered as 
mostly following classical paths, but with a small amount of diffusion from 
one classical path to those nearby in phase-space \cite{Aleiner1996},
where the timescale for this diffusion is set by $t_Q$.  

Consider a quantity which is only non-zero in the vicinity of the $\al$th 
isolated 
periodic path through the system.  If one ignores repetitions of the path
then the integration of that quantity over the surface in phase space with 
energy, $E$, can be written in terms
of coordinates parallel, $(\bi{r}_\pll,\bi{p}_\pll)$, and perpendicular, 
$(\bi{r}_\x,\bi{p}_\x)$, to the path.  Then,
\begin{equation} 
\fl \eqalign{
\int \rmd \bi{r} \ \ (\cdots) 
&=\int \rmd r_\pll \rmd \bi{r}_{\x} \ (\cdots) 
= v_E \int_0^t \rmd t' \int \rmd \bi{r}_{\x} \ (\cdots) \\
\int  \rmd \bi{p}_E \ (\cdots) 
&= \int \rmd p_\pll \rmd \bi{p}_\x \ v_E^{-1} \de 
\sqbrkt{p_\pll-p_E} (\cdots) = v_E^{-1} \int \rmd \bi{p}_\x \ (\cdots)}
\label{1storder170}
\end{equation}
where the $\rmd \bi{p}_E$ is defined just after equation \eref{mid72}.
This was used in \cite{Argaman1993a} to prove \eref{mid70}, 
and will be of great use to us.

The exponent in \eref{1storder110} is in terms of the
two points  $(\bi{r}_1,\bi{p}_1)$ and 
$(\bi{r}_2,\bi{p}_2)$, which are both somewhere on path $\al$. 
It is therefore necessary cast the right-hand side of \eref{1storder120}
in a form which specifies that these points are indeed on path $\al$.  
We will now discuss the details 
of how this is done. Consider a path which starts at $(\bi{r},\bi{p})$. 
We want to know the probability that the path passes through 
$(\bi{r}',\bi{p}')$.  
We do not care when the path passes through $(\bi{r}',\bi{p}')$,
so long as it does so at some time between time $t_a$ and time $t_b$.
So long as the period of the path is more than $t_2$, the probability density 
for this to occur is 
$\int_{t_a}^{t_b} \rmd t' \ 
f_{\al E} \brkt{\bi{r}',\bi{p}',t'; \bi{r},\bi{p}}$.
Therefore the probability of going from $(\bi{r}_0,\bi{p}_0)$ to 
$(\bi{r},\bi{p})$ in time, t, and passing through $(\bi{r}',\bi{p}')$
on the way from $(\bi{r}_0,\bi{p}_0)$ to $(\bi{r},\bi{p})$ is,
\begin{eqnarray}
 \int_0^t \rmd t' \ f_{\al E} \brkt{\bi{r},\bi{p},t-t'; \bi{r}',\bi{p}'} 
f_{\al E} \brkt{\bi{r}',\bi{p}',t'; \bi{r}_0,\bi{p}_0}
\end{eqnarray}
Given that the propagation probability for the path $\al$ from 
$(\bi{r}_0,\bi{p}_0)$ to $(\bi{r},\bi{p})$ in time $t$, is 
$f_{\al E} \brkt{\bi{r},\bi{p},t; \bi{r}_0,\bi{p}_0}$, 
one can {\it insist} that the point $(\bi{r}',\bi{p}')$ is somewhere on 
the path. Then casting all coordinates in terms of components parallel and 
perpendicular to the path,
\begin{eqnarray}
\fl f_{\al E} \brkt{\bi{r},\bi{p},t; \bi{r}_0,\bi{p}_0}
=\int \rmd \bi{r}'_\x \rmd \bi{p}'_\x \nonumber \\
\times \int_0^t \rmd t'
f_{\al E} \brkt{\bi{r},\bi{p},t-t'; \bi{r}',\bi{p}'}
f_{\al E} \brkt{\bi{r}',\bi{p}',t'; \bi{r}_0,\bi{p}_0}
\label{1storder123}
\end{eqnarray}
The integral over all 
$(\bi{r}'_\x,\bi{p}'_\x)$ in the vicinity of the path $\al$ will pick up a 
$\de$-function when $(\bi{r}',\bi{p}')$ is on the path, with no contribution 
the rest of the time.  

As previously mentioned we wish to specify that the points  
$(\bi{r}_1,\bi{p}_1)$ and $(\bi{r}_2,\bi{p}_2)$ are somewhere on path $\al$. 
Using \eref{1storder123} to specify this, \eref{1storder120} becomes,
\begin{eqnarray}
\fl \stbrkt{A_\alpha}^2 \de \brkt{t-T_\alpha} = {t \over (2 \pi \hbar)^2} 
\int_{\Gamma_\alpha} \rmd \bi{r} \rmd \bi{p}_E 
\int \rmd \bi{r}_{1 \x} \rmd \bi{p}_{1 \x} \ 
\rmd \bi{r}_{2 \x} \rmd \bi{p}_{2 \x}  \nonumber \\  \times
\int_0^t \rmd t_1 \int_{t_1}^{t+t_1} \rmd t_2 \ 
f_{\al E} \brkt{\bi{r},\bi{p},t;\bi{r}_2,\bi{p}_2,t_2;\bi{r}_1,\bi{p}_1,t_1;
\bi{r},\bi{p}}
\label{1storder130}
\end{eqnarray}
where the integrals over $\bi{r}_{1,2}$ and $\bi{p}_{(1,2)E}$ are over the 
phase space surface with energy, $E$.
We have defined 
$f_{\al E} \brkt{\bi{r},\bi{p},t;\bi{r}_2,\bi{p}_2,t_2;\bi{r}_1,\bi{p}_1,t_1;
\bi{r},\bi{p}}$ as the 
probability that a classical particle with energy, $E$, which starts at 
$(\bi{r},\bi{p})$, passes through $(\bi{r}_1,\bi{p}_1)$ at time, $t_1$, then 
through $(\bi{r}_2,\bi{p}_2)$ at time, $t_2$, before returning to 
$(\bi{r},\bi{p})$ at time, $t$.  Hence 
\begin{eqnarray}
\fl f_{\al E} \brkt{\bi{r},\bi{p},t;\bi{r}_2,\bi{p}_2,t_2;\bi{r}_1,\bi{p}_1,t_1;
\bi{r},\bi{p}} \nonumber \\ 
\fl \ \ = f_{\al E} \brkt{\bi{r},\bi{p},t-t_2;\bi{r}_2,\bi{p}_2}
f_{\al E} \brkt{\bi{r}_2,\bi{p}_2,t_2-t_1;\bi{r}_1,\bi{p}_1}
f_{\al E} \brkt{\bi{r}_1,\bi{p}_1,t_1;\bi{r},\bi{p}}
\label{1storder140}
\end{eqnarray}
The integrals over $\bi{r}$ and $\bi{p}_E$ are carried out by noting that,
\begin{eqnarray}
\fl \int \rmd \bi{r} \rmd \bi{p}_E \ 
f_{\al E} \brkt{\bi{r},\bi{p},t;\bi{r}_2,\bi{p}_2,t_2;\bi{r}_1,\bi{p}_1,t_1;
\bi{r},\bi{p}} = 
f_{\al E} \brkt{\bi{r}_1,\bi{p}_1,t;\bi{r}_2,\bi{p}_2,t';
\bi{r}_1,\bi{p}_1}
\label{1storder150}
\end{eqnarray}
where we have defined $t'\equiv t_2-t_1$.  The integrand is now independent of 
$t_1$, so the integral over it simply generates a factor of $t$.  
Substituting \eref{1storder150} and \eref{1storder130} into 
\eref{1storder110}, we note that we only expect 
a contribution when $\bi{p}_2$ is approximately anti-parallel to $\bi{p}_1$.
Therefore we can resolve $\bi{r}_2$ and $\bi{p}_2$ in the exponent of 
\eref{1storder110} into
components parallel and perpendicular to the momentum $\bi{p}_1$.  
The integrals over $\bi{r}_{2\x}$ and $\bi{p}_{2\x}$ can then be 
evaluated using the following stationary phase approximation,
\begin{eqnarray}
\int \rmd x \rmd p f(x,p) \exp \sqbrkt{ixp/\hbar} \ \buildrel {\hbar \to 0} 
\over \longrightarrow \ (2 \pi \hbar) f(0,0)
\label{1storder180}
\end{eqnarray}
This means the the only contribution comes from paths where 
$\bi{r}_{2\x} =0$ and $\bi{p}_{2\x} =0$.   
The integral over $\bi{p}_{2\pll}$ is constrained by the energy $\de$-function,
so $\stbrkt{\bi{p}_1} = \stbrkt{\bi{p}_2} = p_E$, and therefore 
$\bi{p}_2 = -\bi{p}_1$.  Hence we find,
\begin{eqnarray}
\fl \stbrkt{A_\al}^2 \de \brkt{t-T_\alpha} 
\exp \sqbrkt{-i \de S_2 /\hbar}
\nonumber \\
= {t^2 \over(2 \pi \hbar)^{3-d}}  
\int \rmd \bi{r}_{1 \x} \rmd \bi{p}_{1 \x} 
\int_0^t \rmd t'
f_{\al E} \brkt{\bi{r}_1,\bi{p}_1,t;\bi{r}_1,-\bi{p}_1,t';\bi{r}_1,\bi{p}_1}
\label{1storder195}
\end{eqnarray}
where $\de S_2$ is given by \eref{mid95}.
The integral over $\bi{r}_{1 \x}$ and $\bi{p}_{1 \x}$
can be converted back to the system's coordinates by noting that  
since the integral over $t'$ makes the right-hand side of 
\eref{1storder195} independent of $r'_\pll$, the result is unchanged 
if we integrate over $r'_\pll$, so long as we divide by the path length, 
$v_E t$.  Then \eref{1storder170} can be used to put the integrals in terms 
of the systems' coordinates.  However when we do this, the integral of 
$\bi{r}_1$ and $\bi{p}_{1E}$ over all phase space double-counts all 
the paths, so we have to divide through by two.  Since we get a double 
contribution from each path if the system has time-reversal symmetry
and no contribution if it does not, we multiply through by $2(2-\be)$.
Hence,
\begin{eqnarray}
\fl \stbrkt{A_\alpha}^2 \de \brkt{t-T_\alpha} 
\exp \sqbrkt{-i \de S_2 /\hbar }
\nonumber \\
={(2-\be)t \over(2 \pi \hbar)^{3-d}} 
\int \rmd \bi{r}_1 \rmd \bi{p}_{1E} \ \int_0^t \rmd t' 
f_{\al E} \brkt{\bi{r}_1,\bi{p}_1,t;\bi{r}_1,-\bi{p}_1,t';\bi{r}_1,\bi{p}_1}
\label{1storder197}
\end{eqnarray}
Finally substituting this into \eref{1storder110}, summing over all 
possible paths, and carrying out the averaging $\anbrkt{\cdots}$, we find,
\begin{eqnarray}
\fl K_2(E,t) = {2(2-\be)t \over(2 \pi \hbar)^{3-d}}  
\int \rmd \bi{r}_1 \rmd \bi{p}_{1E} \ \int_0^t \rmd t' 
\Bigl< f_E \brkt{\bi{r}_1,\bi{p}_1,t;\bi{r}_1,-\bi{p}_1,t';\bi{r}_1,\bi{p}_1}
\Bigr>
\label{1storder200}
\end{eqnarray}
At present we have not specified what form the averaging $\anbrkt{\cdots}$
will take in the definition of the SFF and the TLCF.  
However it is clear from  
\eref{1storder200} that it only affects the 
propagation probability.  So if the averaging were over an energy window near 
the energy, $E$, then one only needs to know what the propagation 
probability is after it has been averaged it over that energy window.  
In this paper we will be applying the above result to a disordered system, 
so we can chose to define the averaging in the definition of the TLCF 
\eref{mid10}, and hence the SFF \eref{mid60}, as 
averaging over an ensemble of systems with different positions of the 
scatterers.  In this case the averaging of the propagation probability in 
\eref{1storder200} would be over this ensemble of systems.

Equation \eref{1storder200} is the two loop term in the weak diagonal 
expansion of the SFF.  When we 
set $\be=1$, it applies to a system with time-reversal symmetry, and is 
the leading order correction to the strong diagonal approximation.  
When we set $\be=2$, it applies to a system without time-reversal 
symmetry, for which the two loop term is zero.  Therefore in the $\be=2$ case, 
it is the three loop term which is the leading order correction to the strong 
diagonal approximation.

\section{The three loop contribution to the weak diagonal expansion 
(for a system without time-reversal symmetry).}

In this paper we are interested in calculating the 
leading order correction to the strong diagonal 
approximation.  As discussed at the beginning of the previous section, the 
two loop term of the weak diagonal perturbation expansion is zero in a 
system without time-reversal symmetry.
Therefore, to find the leading correction to the strong diagonal 
approximation for a system without time-reversal symmetry, it is necessary to 
calculate the three loop term in the expansion. 
In general there are five path geometries that 
contribute to the three loop term in the expansion, however we are only 
interested in the three loop term for a system without time-reversal symmetry. 
In such a system three of these five possible geometries give no 
contribution because one path follows the time reverse of the other at some 
point in their period.  
This means we have the contribution of two geometries
to calculate, these geometries are shown in Fig.\ \ref{fig10}c,d.
 
The first contribution is shown in Fig.\ \ref{fig10}c, the action difference 
between the two paths, $\de S_{3a}$, is given by \eref{mid100}.  
The amplitude, $A_\al$, is given by taking \eref{1storder120} and using 
\eref{1storder123} to insist that $(\bi{r}_1,\bi{p}_1)$, $(\bi{r}_2,\bi{p}_2)$ 
and $(\bi{r}_3,\bi{p}_3)$ are somewhere on the path $\al$,
\begin{eqnarray}
\fl \stbrkt{A_\alpha}^2 \de \brkt{t-T_\alpha} = {t \over (2 \pi \hbar)^2} 
\int_{\Gamma_\alpha} \rmd \bi{r} \rmd \bi{p}_E 
\int \rmd \bi{r}_{1\x} \rmd \bi{p}_{1\x} \ \rmd \bi{r}_{2\x} \rmd \bi{p}_{2\x}
\ \rmd \bi{r}_{3\x} \rmd \bi{p}_{3\x}
\nonumber \\  
\fl \qquad \times
\int_0^t \rmd t_1 \int_{t_1}^{t+t_1}\rmd t_2 \int_{t_2}^{t+t_1}\rmd t_3  \ 
f_{\al E} \brkt{\bi{r},\bi{p},t;\bi{r}_3,\bi{p}_3,t_3;\bi{r}_2,\bi{p}_2,t_2;
\bi{r}_1,\bi{p}_1,t_1;\bi{r},\bi{p}}
\label{2ndorder10}
\end{eqnarray}
The integral over $(\bi{r},\bi{p})$ is carried out using \eref{1storder150}. 
We only expect contributions when $\bi{p}_2$ and $\bi{p}_3$ are 
approximately parallel to $\bi{p}_1$.  
So substituting \eref{mid100} and \eref{2ndorder10} into \eref{mid90}, we
carry out the stationary phase integrals over $(\bi{r}_{2\x},\bi{p}_{2\x})$ 
and $(\bi{r}_{3\x},\bi{p}_{3\x})$, this leaves the integral over 
$(\bi{r}_{1\x},\bi{p}_{1\x})$.  As in the two loop case we introduce an
extra integral over $r_{1\pll}$ and divide through by the length of the path, 
$v_Et$, the integral of $\bi{r}_1$ and $\bi{p}_{1E}$ 
is then over the phase space surface with energy, $E$.
However now this integral over all phase space 
counts each contributing path three times, so we must divide through by 
three.  For completeness we also multiply the result by the time-reversal 
symmetry factor, $2/\be$, so that when $\be=2$ ($\be=1$) we count each 
path once (twice).  However one should note that to calculate the total 
contribution in a system with time-reversal symmetry, one would not
only have to set $\be=1$ in these equations.  One also has to calculate 
the contribution of the trajectories that we ignore because they are zero in 
systems without time-reversal symmetry.
Summing over all paths and carrying out the averaging $\anbrkt{\cdots}$, we 
find,
\begin{eqnarray}
\fl K_{3a}(E,t)={4t \over 3\be (2 \pi \hbar)^{4-2d}} 
\int \rmd \bi{r}_1 \rmd \bi{p}_{1E}  \int_0^t \rmd t'_1 \int_0^{t-t'_1} 
\rmd t'_2 
\nonumber \\ \times
\Bigl<{f_E \brkt{\bi{r}_1,\bi{p}_1,t;\bi{r}_1,\bi{p}_1,t_2';
\bi{r}_1,\bi{p}_1,t_1';\bi{r}_1,\bi{p}_1}}\Bigr>
\label{2ndorder12}
\end{eqnarray}
where we have defined $t'_1 \equiv t_2-t_1$ and $t'_2 \equiv t_3-t_2$.

The second contribution to the three loop term is shown in Fig.\
\ref{fig10}d, the action difference between the two paths is,
\begin{eqnarray}
\de S_{3b} = (\bi{r}_1-\bi{r}_3)\cdot(\bi{p}_3-\bi{p}_1) 
- (\bi{r}_2-\bi{r}_4)\cdot(\bi{p}_4-\bi{p}_2) 
\label{2ndorder20}
\end{eqnarray}
and the amplitude can be written in terms of probabilities,  by taking 
\eref{1storder120} and using \eref{1storder123} to insist that 
$(\bi{r}_1,\bi{p}_1)$, $(\bi{r}_2,\bi{p}_2)$, $(\bi{r}_3,\bi{p}_3)$
and $(\bi{r}_4,\bi{p}_4)$ are somewhere on the path $\al$,
\begin{eqnarray}
\fl \stbrkt{A_\alpha}^2 \de \brkt{t-T_\alpha} = {t \over (2 \pi \hbar)^2} 
\int_{\Gamma_\alpha} \rmd \bi{r} \rmd \bi{p}_E 
\int \rmd \bi{r}_{1\x} \rmd \bi{p}_{1\x} \ \rmd \bi{r}_{2\x} \rmd \bi{p}_{2\x}
\ \rmd \bi{r}_{3\x} \rmd \bi{p}_{3\x} \ \rmd \bi{r}_{4\x} \rmd \bi{p}_{4\x}
\nonumber \\  
\times \int_0^t \rmd t_1  \int_{t_1}^{t+t_1} \rmd t_2 
\int_{t_2}^{t+t_1} \rmd t_3 \int_{t_3}^{t+t_1} \rmd t_4 \nonumber \\  
\times
f_{\al E} \brkt{\bi{r},\bi{p},t;\bi{r}_4,\bi{p}_4,t_4;\bi{r}_3,\bi{p}_3,t_3;
\bi{r}_2,\bi{p}_2,t_2;\bi{r}_1,\bi{p}_1,t_1;\bi{r},\bi{p}}
\label{2ndorder30}
\end{eqnarray}
The integral over $(\bi{r},\bi{p})$ is carried out using \eref{1storder150}. 
We only expect contributions when $\bi{p}_3$ is approximately parallel to 
$\bi{p}_1$, while $\bi{p}_4$ is approximately parallel to $\bi{p}_2$.
Substituting \eref{2ndorder20} and \eref{2ndorder30} into \eref{mid90},  we
carry out the stationary phase integrals over $(\bi{r}_{3\x},\bi{p}_{3\x})$ 
and $(\bi{r}_4,\bi{p}_4)$.  This leaves the integral over 
$(\bi{r}_{1\x},\bi{p}_{1\x})$ and $(\bi{r}_{2\x},\bi{p}_{2\x})$, we
turn these into integrals over the phase space surface with energy, $E$, 
by introducing extra integrals over $r_{1\pll}$ and $r_{2\pll}$, 
while dividing through twice by the path length.
However now the integrals of $\bi{r}_{1,2}$ and $\bi{p}_{(1,2)E}$ over all 
phase space count each contributing path four times, so we must divide 
through by four.  As before, we also multiply the result by the time-reversal 
symmetry factor, $2/\be$.  Summing over all paths and carrying out the 
averaging $\anbrkt{\cdots}$,
\begin{eqnarray}
\fl K_{3b}(E,t)={1 \over \be (2 \pi \hbar)^{4-2d}} 
\int \rmd \bi{r}_1 \rmd \bi{p}_{1E} \ \rmd \bi{r}_2 \rmd \bi{p}_{2E}  
\int_0^t \rmd t'_1 \int_0^{t-t'_1} \rmd t'_2 
\int_0^{t-t'_1-t'_2} \rmd t'_3 \
\nonumber \\ \times
\Bigl<{f_E \brkt{\bi{r}_1,\bi{p}_1,t;\bi{r}_2,\bi{p}_2,t_3';
\bi{r}_1,\bi{p}_1,t_2';\bi{r}_2,\bi{p}_2,t_1';\bi{r}_1,\bi{p}_1}}\Bigr>
\label{2ndorder35}
\end{eqnarray}
where we have defined $t'_1 \equiv t_2-t_1$, $t'_2 \equiv t_3-t_2$ and 
$t'_3 \equiv t_4-t_3$.
The sum of \eref{2ndorder12} and \eref{2ndorder35} gives the three loop term 
in the weak diagonal perturbation expansion of the SFF for a system without 
time reversal symmetry.  Since for such a system the two loop term in the 
expansion is zero, 
this is the leading order correction to the strong diagonal approximation.

%----------------------------------------------
\section{The leading order behaviour of the TLCF in a two-dimensional 
disordered system with time-reversal symmetry.}

We will now apply the results of the weak diagonal expansion to 
a two-dimensional dilute system of
randomly placed semiclassical scatterers.  Each scatterer is semiclassical 
in the sense that its radius, $a$, is much greater than the wavelength, 
$\lambda_F$.  The typical distance between the scatterers is much larger than
their radius.  

In any given system the scatterers are randomly placed.
We will be ensemble averaging our results, where the ensemble to be averaged 
over is an ensemble of systems with 
the same macroscopic properties but with different positions of the scatterers.
The propagation probabilities through two separate regions of the potential 
are uncorrelated because the scatterers are randomly placed, and can 
therefore be ensemble averaged separately.  We will assume,
\begin{eqnarray}
\fl \anbrkt{f_E \brkt{\bi{r}_1,\bi{p}_1,t;\bi{r}_1,-\bi{p}_1,t';
\bi{r}_1,\bi{p}_1}} \nonumber \\
= \anbrkt{f_E \brkt{\bi{r}_1,\bi{p}_1,t-t';\bi{r}_1,-\bi{p}_1}}\
\anbrkt{f_E \brkt{\bi{r}_1,-\bi{p}_1,t';\bi{r}_1,\bi{p}_1}}
\label{1storder210}
\end{eqnarray}
where $\anbrkt{\cdots}$ now denotes ensemble averaging.
This assumption will be valid so long as $t,t' \gg \tErh$, because then the 
region over which the parts of the paths are correlated is a small 
proportion of the whole path.

On these relatively long timescales, the classical behaviour of the system is 
diffusive.  The propagation probability, 
$f_E(\bi{r}',\bi{p}',t;\bi{r},\bi{p})$, is not entirely classical, 
it include some diffractive scatterings, however these are not going to 
change the diffusive nature of the propagation probability distribution.
After a small number of scatterings the direction of the momentum 
of a classical path is effectively randomised.
Therefore the ensemble averaged propagation probability is 
independent of both the initial and final direction of the momentum.  
Hence,
\begin{eqnarray}
\anbrkt{f_E \brkt{\bi{r}',\bi{p}',t;\bi{r},\bi{p}}}= 
W(\bi{r}',\bi{r};t) {v_E \over S_d p_E^{d-1}}
\label{1storder230}
\end{eqnarray}
where $S_d$ is the surface area of a $d$-dimensional unit sphere. 
$W(\bi{r},\bi{r}';t)$ is the probability of a diffusing particle which 
starts at $\bi{r}$ being within $\rmd \bi{r'}$ of $\bi{r'}$ after time, $t$,
it therefore satisfies the diffusion equation,
\begin{eqnarray}
\sqbrkt{{\partial  \over \partial t} - D \nabla_{\bi{r}'}^2} 
W(\bi{r}',\bi{r};t) = \de (t) \de (\bi{r}'-\bi{r})
\end{eqnarray}
For time scales smaller than the ergodic time, $t_{\rm erg} = L^2 / D$, the 
system is in the diffusive regime, the diffusing 
probability distribution $W(\bi{r}',\bi{r};t)$ does not know about the 
boundaries of the container and,
\begin{eqnarray}
W(\bi{r}',\bi{r};t) =W(\bi{r}' \! -\bi{r};t) = (4 \pi D t)^{-d/2} \exp 
\sqbrkt{-\stbrkt{\bi{r}'-\bi{r}}^2 /(4Dt)}
\label{1storder250}
\end{eqnarray}
Disordered systems are believed to obey one-parameter scaling, hence
it is usual to characterise disordered systems in terms of a single parameter, 
their dimensionless conductance, $g$. 
This is simply the conductance measured in units of $e^2/(n_d \hbar)$, where 
the parameter $n_d= \pi (2\pi)^d /S_d$.  The dimensionless conductance of a
disordered system with volume, $L^d$, and diffusion coefficient, $D$, is 
given by,
\begin{equation}
g= n_d E_{Th} / \De
\label{mid30}
\end{equation}
where the Thouless energy, $E_{Th}= \hbar/t_{\rm erg}$.
We will therefore write our results for disordered systems in terms of
the dimensionless conductance. 

If one uses this description of a diffusive system to calculate the strong
diagonal approximation of the SFF \eref{mid70}, one find that 
in two dimensions the result is 
$t$ independent.  Since the TLCF is the Fourier transform of the SFF, 
the strong diagonal approximation of the TLCF is zero.  This means 
that it is the corrections to the strong diagonal approximation that give 
the behaviour of the TLCF in a two dimensional diffusive systems. 
Here we will apply the weak diagonal approximation results derived in the 
previous sections to a diffusive system with time-reversal symmetry.
By using the above description of propagation in a diffusive system, we 
calculate the two loop term in the weak diagonal perturbation expansion for 
such a system.  This will give the leading order behaviour of the TLCF in a
two dimensional diffusive system with time-reversal symmetry. 

We substitute \eref{1storder230} and \eref{1storder200} into \eref{mid50},
and find the semiclassical evaluation of the two loop contribution to the 
TLCF, $R_2^{\rm sc}(\om)$, is given by,
\begin{eqnarray}
\fl R_2^{\rm sc}(\om) = 8(2-\be) L^{2d} \brkt{\De \over 2 \pi \hbar}^3 
\nonumber \\ 
\Re \e \sqbrkt{
\int_0^\infty \rmd t \ t \ \exp \sqbrkt{i \om t /\hbar} \int_0^t \rmd t'
W \brkt{0;t'} W \brkt{0;t-t'}}
\label{1storder240}
\end{eqnarray}
where the level spacing, $\De$, is simply given by 
$\De = h^d v_E / \brkt{S_d p_E^{d-1}L^d}$, where $S_d p_E^{d-1}L^d/v_E$ 
is the volume of the constant energy surface of the phase space of the system.  

Thus substituting \eref{1storder250} into \eref{1storder240},
gives $R_2^{\rm sc}(\om)$ in the diffusive regime, 
\begin{eqnarray}
\fl R_2^{\rm sc}(\om) = {8 (2-\be)L^{2d} \over (4 \pi D)^d} 
\brkt{\De \over 2 \pi \hbar}^3 \nonumber \\ 
\Re \e \sqbrkt{
\int_0^\infty \rmd t \ t \ \exp \sqbrkt{i \om t /\hbar} \int_0^t \rmd t'
t'^{-d/2} \brkt{t-t'}^{-d/2} }
\end{eqnarray}
One can evaluate the integral over $t'$ by noting it is a Euler 
$\beta$-function and so can be  written in terms of $\Gamma$-functions.
The result in 2 dimensions can be found by setting $d=2+\epsilon$, and 
expanding the $\Gamma$-functions in $\epsilon$.

However to compare the results of the semiclassics with those of the
impurity diagram technique, we can Fourier transform the ``$W$''s
in \eref{1storder240} and then compare directly the functional forms of 
the results in the two approaches.
If we define $W \brkt{\bi{q},\om}$ using,  
\begin{eqnarray}
W \brkt{\bi{r};t} = 
\int {{\rmd}^d \bi{q} \ \rmd \om \over \brkt{2 \pi \hbar}^{d+1}}
\exp \sqbrkt{-i \brkt{\bi{q} \cdot \bi{r}+\om t}/ \hbar}
W \brkt{\bi{q},\om} 
\label{1storder260}
\end{eqnarray}
then in the diffusive regime, 
\begin{eqnarray}
W \brkt{\bi{q},\om} = \hbar^2 \sqbrkt{D \bi{q}^2 -i\hbar \om}^{-1}
\label{1storder265}
\end{eqnarray}
Substituting \eref{1storder260} and \eref{1storder265} into 
\eref{1storder240}, and carrying out the
integrals over $t$ and $t'$, we find 
the semiclassical result for the two loop term in the 
weak diagonal perturbation expansion,
\begin{eqnarray}
R_2^{\rm sc}(\om) =  8(2-\be) \hbar L^{2d} \brkt{\De \over 2 \pi \hbar}^3 
\Re \e \sqbrkt{
{\partial \sqbrkt{I_1(\om)}^2\over \partial (i\om)} }
\label{1storder270}
\end{eqnarray}
in deriving this we have defined,
\begin{equation}
\eqalign{
I_n(\om) &\equiv \int {{\rmd}^d \bi{q} \over \brkt{2 \pi \hbar}^d} \ 
\hbar^{2n} \sqbrkt{D \bi{q}^2 -i\hbar \om}^{-n} \\
&= {\Gamma(n- d/2) \over (4 \pi)^{d/2}(n-1)!} \brkt{\hbar \over \De}^n 
\brkt{n_d \over L^2g}^{d/2} \brkt{-i{\om \over \De}}^{d/2-n}
}
\label{1storder275}
\end{equation}
where $g$ is the dimensionless conductance given by \eref{mid30},
and $\De$ is the average level spacing. 

\begin{figure}
\epsfxsize=0.9\textwidth
\centerline{\epsfbox{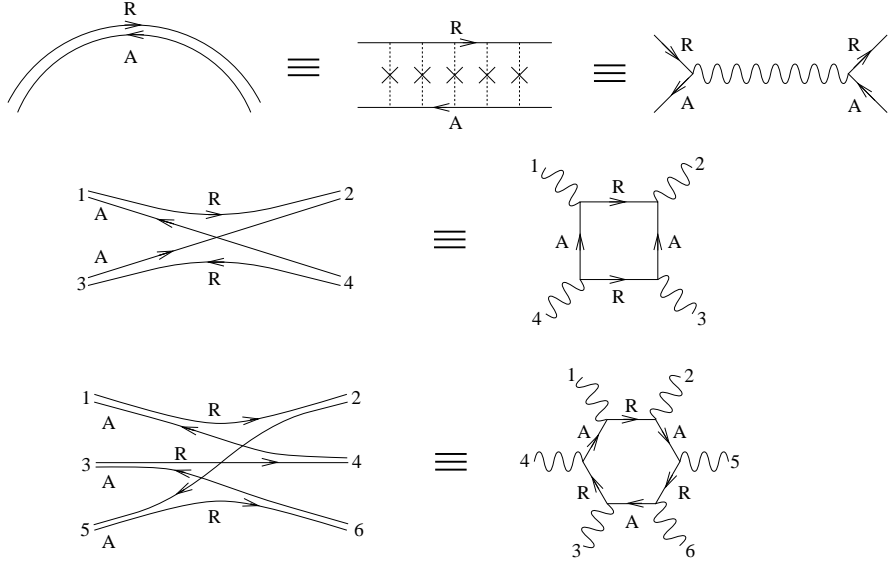}}
\caption{This dictionary allowing a translation from the trajectory
pictures to the diagrams in the impurity diagram technique.
A region where the two paths are identical corresponds to a diffuson (or 
a Cooperon, if one of the arrows is reversed).  The semiclassical Hikami 
boxes are shown with the corresponding Hikami boxes in the impurity diagram 
technique. }
\label{fig30}
\vspace*{0.1in}
\end{figure}

We will now calculate the equivalent term in the impurity diagram technique 
\cite{Abrikosov1969,sigma, Hik:81}.
The trajectory pictures used in the semiclassical approach (see Fig.\ 
\ref{fig10} and \ref{fig20}) can easily be mapped onto the diagrams in the
impurity diagram technique \cite{Smith1998}.  
A dictionary for converting between the two is 
shown in Fig.\ \ref{fig30}.  However this correspondence is fairly 
superficial, the picture drawn in the two approaches maybe similar, but the 
approach to calculations is different.
Two dimensions is the critical dimensionality for the impurity diagram 
technique \cite{Wegner1979}.  As a result renormalisation ideas apply, and 
one can ignore short trajectories and short range behaviour 
(in other words the large $\om$ and $\bi{q}$ behaviour).  By ignoring these
ultraviolet divergences, the impurity diagram technique calculations are 
greatly simplified.

%%%%%%%%%%%%%%%%%%%%%%%%%%%%%%%%%%%%%%%%%%%%
%%%%%% figure 3 (whitfig30.eps) goes here %%%%%%
%%%%%%%%%%%%%%%%%%%%%%%%%%%%%%%%%%%%%%%%%%%%
%%%%%%%%%%%%%%%%%%%%%%%%%%%%%%%%%%%%%%%%%%%%

The two loop term in the impurity diagram technique is given by the 
trajectories shown in Fig.\ \ref{fig10}b, which can be converted to the more 
usual diagrammatics picture using Fig.\ \ref{fig30}.
After throwing away the ultraviolet divergences, the two loop contribution is,
\begin{eqnarray}
R_2(\om) = 4(2-\be) \hbar L^{2d} \brkt{\De \over 2 \pi \hbar}^3 
\Re \e \sqbrkt{
{\partial^2 \over \partial (i\om)^2} \sqbrkt{i\om (I_1(\om))^2}}
\label{1storder279}
\end{eqnarray}
carrying out one of the differentials over $\om$,
\begin{eqnarray}
\fl R_2(\om) =  4(2-\be) \hbar L^{2d} \brkt{\De \over 2 \pi \hbar}^3 
\Re \e \sqbrkt{{\partial \over \partial (i\om)} 
\sqbrkt{(I_1(\om))^2 + {2i \om \over \hbar} I_1(\om)I_2(\om)}}
\label{1storder280}
\end{eqnarray}
Since the divergent terms can only be ignored in the vicinity of the critical 
dimension, $d=2$, one must remember that \eref{1storder279} and hence 
\eref{1storder280}, are only true for $d=2+\epsilon$ where $\epsilon$ is small.

The functional form of the results of the two approaches, 
\eref{1storder270} and \eref{1storder280}, are different.  The 
impurity diagram technique result has an extra term, however it so happens 
that in two dimensions, this term is purely imaginary and so does not 
contribute to  the TLCF.  
This is most easily shown by evaluating its contribution in
$d=2+\epsilon$, writing the integrals in terms of $\Gamma$-functions and then
taking $\epsilon$ to zero.
Therefore in two dimensions the two approaches have the same 
functional form.  In two dimensions the semiclassical result simply disagrees 
with the impurity diagram result by a factor of two. 
Since the impurity diagram 
technique result \eref{1storder280} is only true for two dimensions, our 
conclusion is that the semiclassics gets the right answer but with the 
prefactor wrong by a factor of two.  However it looks suspiciously like 
it is so close to the correct result only as the result of a lucky cancellation 
of the term in the impurity diagram technique that the semiclassics can not 
reproduce.  To confirm if this is the 
case we need to go to the next order in the weak diagonal expansion, and 
compare that result to the three loop term in the impurity diagram technique.

For the sake of completeness we now evaluate $R_2^{\rm sc}(\om)$ in $d=2$,
taking \eref{1storder270} setting $d=2+\epsilon$ where $\epsilon$ is small,
\begin{eqnarray}
R_2^{\rm sc}(s\De) = -{(2-\be) \over \pi g^{2+\epsilon}} {1 \over \epsilon}
\Re \e \sqbrkt{(-is)^{\epsilon-1}}
\ \buildrel {\epsilon \to 0} \over \longrightarrow \ 
-{(2 - \be) \over 2g^2 |s|}
\label{1storder290}
\end{eqnarray}
where $I_1(\om)$ was evaluated using \eref{1storder275}, 
$g$ is the dimensionless conductance given by \eref{mid30}, 
$s= \om / \De$, $\be=1$ ($\be=2$) for a system with (without) time-reversal 
symmetry.  As we have stated, this result is a factor of two larger than the 
equivalent impurity diagram result.

\section{The leading order behaviour of the TLCF in a two-dimensional 
disordered system without time-reversal symmetry.}

We now evaluate the first of the two semiclassical contributions to the three 
loop term in the weak diagonal expansion of the TLCF, $R_{3a}^{\rm sc}(\om)$, 
for a disordered system without time-reversal symmetry in the diffusive 
regime.  Substituting \eref{1storder210} and \eref{1storder230} into 
\eref{2ndorder12}, we find,
\begin{eqnarray}
\fl R_{3a}^{\rm sc}(\om)=
{16 L^{3d} \over 3\be}\brkt{\De \over 2 \pi \hbar}^4
\Re \e \Bigg[
\int_0^\infty \rmd t \ t \ \exp \sqbrkt{i \om t /\hbar} \nonumber \\ \times
\int_0^t \rmd t'_1 \rmd t'_2 
W \brkt{0;t'_1} W \brkt{0;t'_2} W \brkt{0;t-t'_1-t'_2} \Bigg]
\label{2ndorder15}
\end{eqnarray}
In the diffusive regime, one can use \eref{1storder250} to show this 
becomes,
\begin{eqnarray}
\fl R_{3a}^{\rm sc}(\om)=
{16 L^{3d} \over 3\be(4 \pi D)^{3d/2}}\brkt{\De \over 2 \pi \hbar}^4
\Re \e \Bigg[
\int_0^\infty \rmd t \ t \ \exp \sqbrkt{i \om t /\hbar} \nonumber \\ \times
\int_0^t \rmd t'_1 \rmd t'_2 
\Bigl[t'_1 t'_2 \brkt{t-t'_2-t'_1}\Bigr]^{-d/2}\Bigg]
\end{eqnarray}
However as before we Fourier transform \eref{2ndorder15} to the 
$(\bi{q},\om)$-representation to allow comparison with the impurity diagram 
technique.  Substituting \eref{1storder260} and \eref{1storder265} into 
\eref{2ndorder15} and carrying out the
integrals over $t$ and $t'_{1,2}$, 
\begin{eqnarray}
R_{3a}^{\rm sc}(\om)=
{16 \hbar L^{3d} \over 3\be}\brkt{\De \over 2 \pi \hbar}^4
\Re \e \sqbrkt{{\partial \ \over \partial (i\om)} \sqbrkt{I_1(\om)}^3}
\label{2ndorder16}
\end{eqnarray}
In $(2+\epsilon)$ dimensions where $\epsilon$ is small, this becomes, 
\begin{eqnarray}
R_{3a}^{\rm sc}(s\De)={1 \over 6 \pi \be g^3}{1 \over \epsilon^2} 
\Re \e\sqbrkt{\brkt{-is}^{(3 \epsilon /2) -1}}
\label{2ndorder16.5}
\end{eqnarray}

We now evaluate the second of the two semiclassical contributions, 
$R_{3b}^{\rm sc}(\om)$, for a 
disordered system without time-reversal symmetry in the diffusive regime.  
Substituting \eref{1storder210} 
and \eref{1storder230} into \eref{2ndorder35}, we find, 
\begin{eqnarray}
\fl R_{3b}^{\rm sc}(\om)=
{4 L^{3d} \over \be}\brkt{\De \over 2 \pi \hbar}^4
\Re \e \Bigg[ \int \rmd \bi{r}  
\nonumber \\ \times
\int_0^\infty \rmd t \ t \ \exp \sqbrkt{i \om t /\hbar}
\int_0^t {\rmd t'_1 \over t} \int_0^{t-t'_1} \rmd t'_2 
\int_0^{t-t'_1-t'_2} \rmd t'_3  \
\nonumber \\ \times
W \brkt{\bi{r};t'_1} W \brkt{-\bi{r};t'_2} W \brkt{\bi{r};t'_3}
W \brkt{-\bi{r};t-t'_1-t'_2-t'_3} \Bigg]
\label{2ndorder40}
\end{eqnarray}
where $\bi{r}=\bi{r}_2-\bi{r}_1$.  Substituting in \eref{1storder250} 
this becomes,
\begin{eqnarray}
\fl R_{3b}^{\rm sc}(\om)=
{2 L^{3d} \over \be(4 \pi D)^{2d}}\brkt{\De \over 2 \pi \hbar}^4
\Re \e \Bigg[ \int \rmd \bi{r}  
\int_0^\infty \rmd t \ t \ \exp \sqbrkt{i \om t /\hbar}
\nonumber \\ \times
\int_0^t {\rmd t'_1 \over t} \int_0^{t-t'_1} \rmd t'_2 
\int_0^{t-t'_1-t'_2} \rmd t'_3  \
\Bigl[t'_1 t'_2 t'_3 \brkt{t-t'_3-t'_2-t'_1}\Bigr]^{-d/2} 
\nonumber \\ \times
\exp \sqbrkt{-{\stbrkt{\bi{r}}^2 \over 4D} \brkt{ {1 \over t'_1} +
{1 \over t'_2} + {1 \over t'_3} + {1 \over t-t'_3-t'_2-t'_1}}} \Bigg]
\end{eqnarray}
However we proceed by Fourier transforming \eref{2ndorder40} to the 
$(\bi{q},\om)$-representation to allow direct comparison with the impurity 
diagram technique. 
Substituting \eref{1storder260} into \eref{2ndorder40} and carrying out the
integrals over $\bi{r}$, $t$ and $t'_{1,2,3}$.  
\begin{eqnarray}
\fl R_{3b}^{\rm sc}(\om) \propto
\Re \e \Bigg[ \brkt{e^{i\pi/2}}^{3d/2-4} 
\int \rmd^d \bi{q}_1 \rmd^d \bi{q}_2 \rmd^d \bi{q}_3
\nonumber \\ \times
\sqbrkt{\brkt{\bi{q}_1^2 +1}\brkt{\bi{q}_2^2 +1}\brkt{\bi{q}_3^2 +1}
\brkt{\brkt{\bi{q}_1+\bi{q}_2+\bi{q}_3}^2 +1}}^{-1}
\Bigg]
\end{eqnarray}
For $d=2$, the integral is convergent, so the term in the square 
bracket has no real part, and hence $R_{3b}^{\rm sc}(\om)=0$.

Since the semiclassical result for $R_{3b}^{\rm sc}$ is 
zero, the total semiclassical result
for the three loop term of the expansion, in a $(2+\epsilon)$-dimensional 
disordered system without time-reversal symmetry, is given by
\eref{2ndorder16.5} with $\be=2$.  
Expanding the in powers of $\epsilon$, and taking the limit $\epsilon \to 0$,
we neglect the unphysical $1/\epsilon$ divergent term.
Then in 2-dimensions,
\begin{eqnarray}
R_3^{\rm sc} (s\De)={3 \over 32 g^3} {\ln |s| \over |s|}
\label{2ndorder80}
\end{eqnarray}
where $s=\om / \De$.
We now compare this result with the impurity diagram result for $R_3(\om)$,
in a system without time reversal symmetry ($\be=2$).
The geometries of paths that contribute to $R_3(\om)$ are shown in Fig.\ 
\ref{fig10}c,d, these can be put in terms of the usual impurity diagrams 
using Fig.\ \ref{fig30}.  
In dimension $d=2+\epsilon$ where $\epsilon$ is small, one finds,
\begin{eqnarray}
R_{3}(s\De) = - {\hbar L^{3d}} \brkt{\De \over 2 \pi \hbar}^4 
\brkt{\third \epsilon - \half \epsilon^2}
\Re \e \sqbrkt{{\partial^2 \over \partial (i\om)^2} 
\sqbrkt{i\om (I_1(\om))^3}}
\label{2ndorder90}
\end{eqnarray}
Using the result for $I_1(\om)$ given in \eref{1storder275}, this becomes, 
\begin{equation}
R_{3a}(s\De) = -{1 \over 32\pi g^3} 
{1- \textstyle{\frac{9}{4}} \epsilon^2 \over \epsilon}
\Re \e \sqbrkt{(-is)^{3\epsilon/2-1}}
\buildrel {\epsilon \to 0} \over \longrightarrow \
-{3 \over 128g^3 |s|} 
\label{2ndorder100}
\end{equation}
This result has a $\brkt{-|s|^{-1}}$ dependence, while the semiclassical 
result 
has a $\brkt{\ln |s|/|s|}$ dependence.  Therefore the semiclassical approach 
gives a result which is at odds with the impurity diagram result.

We find it instructive to compare the contributions of individual geometries
in the two approaches.  The physically important quantity is $R_3(\omega)$,
and one would not necessarily expect $R_{3a}(\omega)$ and 
$R_{3b}(\omega)$ to be the same in the two approaches.  However it is
of interest to see where the difference between the two approaches  
comes about.  
Firstly we will consider the impurity diagram result for $R_{3a}(\omega)$,
this result is the contribution of the path geometry shown in Fig.\ 
\ref{fig20}c,
\begin{eqnarray}
R_{3a}(\om) = {8 \hbar L^{3d} \over 3\be} \brkt{\De \over 2 \pi \hbar}^4 
\Re \e \sqbrkt{
{\partial^2 \over \partial (i\om)^2} \sqbrkt{-{3 i\om\over 2} (I_1(\om))^3}}
\end{eqnarray}
carrying out one of the differentials over $\om$,
\begin{eqnarray}
\fl R_{3a}(\om) = - {4 \over \be} \hbar L^{3d} \brkt{\De \over 2 \pi \hbar}^4 
\Re \e \sqbrkt{
{\partial \over \partial (i\om)} 
\sqbrkt{(I_1(\om))^3+ {3i \om \over \hbar} (I_1(\om))^2I_2(\om)}}
\label{2ndorder17}
\end{eqnarray}
In $(2+\epsilon)$-dimensions where $\epsilon$ is small, this becomes, 
\begin{eqnarray}
R_{3a}(s\De)={3 \over 8 \pi \be g^3}{1 + \textstyle{\frac{3}{2}}\epsilon \over 
\epsilon^2} 
\Re \e\sqbrkt{\brkt{-is}^{(3 \epsilon /2) -1}}
\label{2ndorder17.5}
\end{eqnarray}
Comparing \eref{2ndorder16} and \eref{2ndorder17}, we see that the impurity 
diagram approach has an extra 
term not present in the semiclassical result, just as in the previous order.  
Like in the two loop case the semiclassics gets the wrong prefactor on the 
term it does reproduce, in this case a factor of $-3/4$.
However much worse than this, the second term in \eref{2ndorder17} has a 
finite real part in two dimensions, unlike the second term in the two loop 
diagrammatic result \eref{1storder280}.  This means the
semiclassical result $R^{\rm sc}_{3a}$ is very different from the 
diagrammatic result for $R_{3a}$.

We now consider the impurity diagram result for $R_{3b}(\omega)$ 
in dimension $d=2+\epsilon$, where $\epsilon$ is small.  We find,
\begin{equation}
\eqalign{
\fl R_{3b}(\om) &= - {2\hbar L^{3d}\over \be} \brkt{\De \over 2 \pi \hbar}^4 
\brkt{-2 + \third \epsilon - \half \epsilon^2}
\Re \e \sqbrkt{{\partial^2 \over \partial (i\om)^2} 
\sqbrkt{i\om (I_1(\om))^3}}\\
\fl &={3 \over 16 \be \pi} 
{2 + \textstyle{\frac{8}{3}}\epsilon +\textstyle{\frac{3}{4}}\epsilon^3 
\over \epsilon^2}
{1 \over g^3} \Re \e \sqbrkt{(-is)^{3\epsilon/2-1}}
}
\label{2ndorder60}
\end{equation}
where $I_1(\om)$ was evaluated using \eref{1storder275}, 
$g$ is the dimensionless conductance given by \eref{mid30}, 
and $s= \om / \De$.  In the limit that $\epsilon \to 0$ it is clear that 
\eref{2ndorder60} is not zero, therefore we find that the semiclassical and 
impurity diagram technique results are completely different.

The difference between the results of the two approaches can be summarised
as follows.
In the impurity diagram technique there is no $\brkt{\ln |s|/|s|}$ term 
because
the terms in \eref{2ndorder17.5} and \eref{2ndorder60} which are of order
$\brkt{\epsilon^{-2}(-is)^{3\epsilon/2-1}}$ exactly cancel each other.  
In the semiclassics there is no cancellation, and therefore the presence of a 
prefactor of order $\brkt{\epsilon^{-2}}$ means one must expand 
$\brkt{(-is)^{3\epsilon/2-1}}$ to order $\brkt{\epsilon^2}$ to get the 
$\epsilon$-independent term.  It is this that generates the 
$\brkt{\ln |s|/|s|}$ term.

%---------------------------------------------
\section {The weak diagonal expansion for a quantum chaotic system in the 
ergodic regime.} 

It is straightforward to apply the results of the weak diagonal expansion 
to a system in the ergodic regime.  If a system is ergodic, 
a particle is equally likely to be found anywhere in the system's phase 
space.  This means that the propagation probability for a particle 
in such a system is featureless and hence the same for 
all ergodic systems.  The method of evaluating the weak diagonal expansion 
presented here is cast purely in terms of this propagation probability
and therefore predicts the same level statistics for all chaotic and 
disordered systems in the ergodic regime.  
Given this statement one would like to know two things:  (i) are the results 
of the  method discussed in this paper to be believed when applied to the 
ergodic regime, (ii) are the level statistics predicted by this 
method the same as those of RMT?
A partial answer to the first question would be that it would be difficult 
to believe the results of the method in the ergodic regime, considering the 
failure of the method in the diffusive regime of a disordered system.
However we can present a better answer to the first question by considering 
the answer to the second question.  
While the conjecture that chaotic systems have RMT level
statistics \cite{Bohigas1984b} is generally considered unproven,
this is not the case for disordered systems, for which a proof exists 
\cite{Efetov1982+1997}.
Therefore if this method's prediction for the ergodic regime of a disordered 
system does not agree with RMT \cite{Mehta1991}, 
then this can be considered 
a proof that the method does not work in the ergodic regime.

To evaluate the weak diagonal approximation results in the ergodic regime, 
we simply note that in this regime, the average propagation probability, 
$\anbrkt{f_E \brkt{\bi{r}',\bi{p}',t;\bi{r},\bi{p}}}$ is just the inverse 
volume of the phase space surface with energy, $E$.   Thus for a system of
volume $L^d$, containing semiclassical scatterers, the propagation 
probability is, 
\begin{eqnarray}
\anbrkt{f_E \brkt{\bi{r}',\bi{p}',t;\bi{r},\bi{p}}}= 
 {v_E \over L^d S_d p_E^{d-1}}
\label{ergod10}
\end{eqnarray}
We can now construct an expansion for the SFF, $K(t)$, in the small parameter
$t /t_{\rm H}$.  
This expansion can then be compared to the small $t$ expansion of the SFF 
for random matrices \cite{Mehta1991}.  If each term in the weak diagonal 
expansion does not agree with the corresponding term in the expansion for 
random matrices, then we have proven that this method of evaluating the 
weak diagonal approximation is wrong.

The first term in the SFF is linear in $t /t_{\rm H}$ and is given by 
the strong diagonal approximation which agrees with 
the leading order RMT result \cite{Berry1985}. 
It is well known that for $\beta\ne2$ RMT predicts corrections
in powers of $t /t_{\rm H}$, while for $\beta=2$ the linear in $t$ behaviour
is exact for $t<t_{\rm H}$. The weak diagonal 
expansion, if correct, must reproduce the higher-order contributions
 for $\beta=1$ and yet yield no higher-order contributions  for $\beta=2$.

In the ergodic regime with time-reversal symmetry ($\beta=1$)
 the leading  weak-diagonal contribution is the 
two loop term given by substituting \eref{ergod10} into \eref{1storder200}.
Comparing this result with the $\beta=1$ RMT result,
we immediately see it is positive, while the RMT result is 
negative.  Hence we find that this method of evaluating the weak diagonal 
approximation predicts level statistics different from those of 
GOE random matrices.  

In the $\beta=2$ case, the leading order 
 weak diagonal corrections are given by  the three loop terms.
It is well known from the impurity diagram technique that in disordered 
systems such contributions are 
individually non-zero, but that at each order in the 
perturbation expansion they cancel each other.
However when we calculate the three loop terms
 from the semiclassical approach, we find that both $K_{3a}$ and 
$K_{3b}$ (Fig.\ \ref{fig10}c,d) 
are positive so there is no possibility of them cancelling each 
other.  This means that again this method fails to reproduce
the RMT results. 

We reiterate that, since we know that disordered systems have random matrix 
level statistics, the results of this evaluation of the weak diagonal 
approximation must be wrong.

\section{Concluding remarks.}

The main aim of this paper has been to cast the 
weak diagonal expansion in terms of semiclassics so that it could 
be applied to a generic chaotic system.  
If this could be achieved, the perturbation expansion would be a systematic 
way of finding corrections to the Berry diagonal approximation;
thus in disordered systems such corrections 
give rise to weak localisation phenomena.  
If level statistics  of chaotic systems are to be RMT, 
these corrections must be present to describe the difference between
GOE statistics and the Berry diagonal approximation for times much shorter
than the Heisenberg time. 

Having presented this general scheme, we then developed a method of 
evaluating the terms in the weak diagonal expansion based on 
the Gutzwiller trace formula.
Unfortunately, this method could not correctly reproduce the Hikami boxes
which are regions in the phase space where quantum scattering is important.   
Thus it could neither reproduce the weak localisation expansion 
for disordered systems, nor give the corrections to 
the Berry diagonal approximation predicted by RMT. 
We do not believe that any existing semiclassical 
approach is capable of correctly evaluating the Hikami boxes for a 
generic chaotic system.  A method exists \cite{Aleiner1996} in which Hikami 
boxes were calculated for the system of semiclassical scatterers considered 
here.  The problem with this method is that there is no obvious way to 
generalise it to generic chaotic systems.

Finally we wish to encourage the use of disordered systems as a test bed for 
theories of quantum chaos.  It is one of the few examples of chaotic 
systems whose quantum behaviour is well understood.  In particular, any theory
which claims to explain the RMT behaviour of quantum chaotic systems
 should be tested by checking if it produces the correct level 
statistics for disordered systems in the diffusive regime.  It also appears
that it is not enough for a theory to get the correct weak localisation 
correction to the conductivity of a diffusive system --
this is too simple a test to be trusted.
 A complete semiclassical theory
must be able to correctly reproduce higher order corrections to the 
conductivity and the TLCF.

%---------------------------------------------
\ack

We are grateful to Oded Agam and Jon Keating for useful discussions.
I.V.L thanks the EPSRC for partial support under grant GR/K95505.

\null
%---------------------------------------------

\vfill \eject
%---------------------------------------------
\end{document}